# Artificial Tongue-Placed Tactile Biofeedback for perceptual supplementation: application to human disability and biomedical engineering

N. Vuillerme, O. Chenu, A. Moreau-Gaudry, J. Demongeot and Y. Payan

*Abstract* — The present paper aims at introducing the innovative technologies, based on the concept of "sensory substitution" or "perceptual supplementation", we are developing in the fields of human disability and biomedical engineering. Precisely, our goal is to design, develop and validate practical assistive biomedical and/technical devices and/or rehabilitating procedures for persons with disabilities, using artificial tongue-placed tactile biofeedback systems.

Proposed applications are dealing with: (1) pressure sores prevention in case of spinal cord injuries (persons with paraplegia, or tetraplegia); (2) ankle proprioceptive acuity improvement for driving assistance in older and/or disabled adults; and (3) balance control improvement to prevent fall in older and/or disabled adults.

This paper presents results of three feasibility studies performed on young healthy adults.

## I. INTRODUCTION

The concept of "sensory substitution" was introduced and extensively studied by Paul Bach-y-Rita and colleagues in the context of tactile visual substitution systems [1,2]. These researchers evidenced that stimulus characteristics of one sensory modality (e.g., a visual stimulus) could be transformed into stimulations of another sensory modality (e.g., a tactual stimulation). Claiming that "we see with the brain, not the eyes" [1,2], Bach-y-Rita gives the example of a blind individual, without any pulse train coming form the retina, who can compensate and substitute with another modality: this person is able to navigate with a long cane and has a 3D perception of a room and/or a step. Here, whereas the interaction between the cane and the body is a tactile sensation in the hand, the blind person perceives a clear mental image of his 3D environment. Following the same idea, Bach-y-Rita has proposed tactile vision substitution systems to provide visual information to the brain through arrays of mechanical or electrical stimulators in contact with the abdomen, the chest, the brow, the back, the finger or the tongue. Optical images are recorded by a TV camera and transduced (through subsampling of the image) into vibratory or electrical stimulations that are mediated by the skin receptors. After sufficient training, subjects perceive an image in space rather than a tactile stimulation [3,4]. Moreover, subjects fulfilled a shape-recognition task and even experienced a "projection" of the objects they tactually perceived in the external world [2]. For instance, a sudden change in the zoom of the camera caused subjects to act as if they were approaching an obstacle. While the first tactile visual substitution systems were composed of 400 stimulators (20×20 matrix, Ø 1mm each) placed on different skin region of the human body, Bach-y-Rita recently converged to the electro-stimulation of tongue surface [5], which overcomes the practical problems posed by the mechanical stimulation of the skin. Indeed, the human tongue is a highly dense [6], sensitive and discriminative (spatial threshold = 2 mm) array of tactile receptors that are similar to the ones of the skin. Moreover, the high conductivity offered by the saliva insures a highly efficient electrical contact between the electrodes and the tongue surface and therefore does not require high voltage and current [5]. Along these lines, a practical human-machine interface, the *Tongue Display Unit* (TDU) [5], was developed and recently evaluated for blind persons [4]. It consists in a 2D array of miniature electrodes (12×12 matrix) held between the lips and positioned in close contact with the anterior-superior surface of the tongue. A flexible cable connects the matrix to an external electronic device delivering the electrical signals that individually activate the electrodes and therefore the tactile receptors of the tongue.

The present paper aims at introducing the innovative technologies, based on the concept of "sensory substitution" [1,2] or "perceptual supplementation" [7], we are developing in the fields of human disability and biomedical engineering. Precisely, our goal is to design, develop and validate practical assistive biomedical and/or technical devices and/or rehabilitating procedures for persons with disabilities, using artificial tongue-placed tactile biofeedback systems.

Proposed applications are dealing with: (1) pressure sores prevention in case of spinal cord injuries (persons with

Nicolas Vuillerme is with the TIMC-IMAG Laboratory, UMR CNRS 5525, Faculté de Médecine de Grenoble, Bâtiment Jean Roget, F38706 La Tronche Cédex; phone: +33 4 76 63 74 86; fax: +33 4 76 51 86 67; mail: Nicolas.Vuillerme@imag.fr

Olivier Chenu, Alexandre Moreau-Gaudry, Jacques Demongeot and Yohan Payan are also with the TIMC-IMAG Laboratory, UMR CNRS 5525, La Tronche, France. Mail: firstname.lastname@imag.fr

paraplegia, or tetraplegia); (2) ankle proprioceptive acuity improvement for driving assistance in older and/or disabled adults; and (3) balance control improvement to prevent fall in older and/or disabled adults.

For each of these applications, a home-made TDU, consisting of a 6×6 matrix of electrodes (Figure 1A), thus reducing by a factor of 4 the overall dimension of the Bach-y-Rita's device [5], was used. On the one hand, indeed, the addressed biomedical applications do not require the dense 12×12 TDU resolution of the tactile visual substitution systems. On the other hand, the size reduction allowed us to develop a wireless tongue placed tactile biofeedback device that will make its daily use comfortable from an ergonomic point of view (see Discussion section).

The following sections present three feasibility experiments performed on young healthy adults.

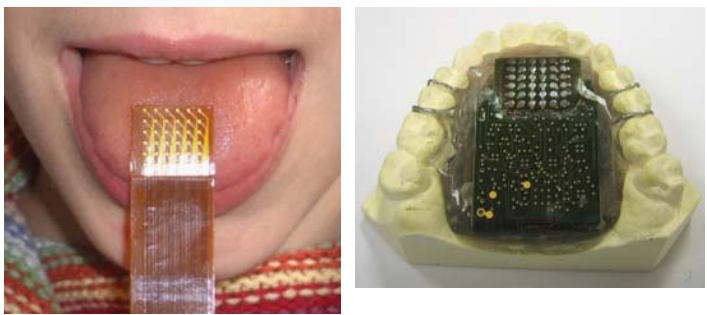

**Figure 1A**: Wire TDU: 6×6 matrix and flexible cable

**Figure 1B**: Wireless TDU: 6×6 matrix and embedded radio-frequency circuit

## II. PRESSURE SORES PREVENTION

**Objectives**

A pressure sore is defined as an area of localized damage to the skin and underlying tissue caused by overpressure, shearing, friction or a combination of these factors. Its prevalence ranges from 23% to 39% in adults with spinal cord injuries [8,9] since this population do not get the "signal" arising from the buttock area that allows healthy subjects to prevent pressure sores by moving their body in a conscious or subconscious manner. Located near bony prominences such as the ischium, sacrum and trochanter, pressure sores are recognized as the main cause of rehospitalization for patients with paraplegia [10]. Their treatment, which can be medical or surgical, is always long, difficult and expensive.

Within this context, we developed an original system for preventing the formation of pressure sores in individuals with paraplegia. Its underlying principle consists in (1) putting onto the wheelchair seat area a pressure mapping system that allows real-time acquisition of the pressure applied on the seat/skin interface ; and (2) sending a "signal" to the individual through the TDU matrix each time an overpressure zone is detected by the pressure mapping system. This signal was voluntary chosen as a simple low-level message: if the overpressure is supposed to disappear with a postural change of the patient body in the front (respectively back, left and right) direction, the corresponding six electrodes of the front (respectively back, left and right) row of the matrix are activated (Figure2).

The purpose of the present experiment was to assess the performance of this system in young healthy adults.

**Methods**

Subjects

Ten young healthy university students (mean age: 26.2 years) were included in this study. They gave their informed consent to the experimental procedure as required by the Helsinki declaration (1964) and the local Ethics Committee, and were naive as to the purpose of the experiment.

Task and procedures

Subjects were seated comfortably in a chair. Electrostimulation of one of the four rows of the matrix (figure 2) were sent to the TDU and subjects were asked to move their chest according to the felt electro-stimulated direction. After each movement following the electro-stimulated direction, a new record of the pressure map was provided.

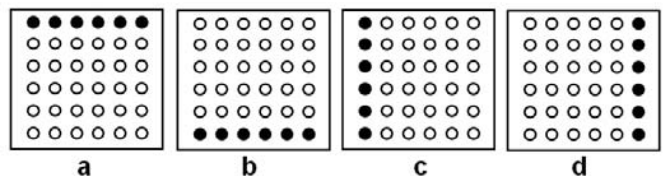

**Figure 2.** Different patterns for electro-stimulation.

When the six electrodes placed in front of the electrode array were activated, subjects were asked to move their chest forward (a). In the same way, they were asked to move their chest backward, to the left or to the right if electrodes situated behind (b), on the left (c) and on the right (d) were activated.

Data analysis

The pressure map applied at the seat/skin interface was recorded at a frequency of 10Hz. By computing the differences between the two pressure maps (before *versus* after electro-stimulation), we determined whether or not the movement was adapted to the electro-stimulated information. In the first case, the result was marked as "one", and otherwise, as "zero". For each subject, the experiment was carried out 10 times, thus obtaining a total score out of 10.

**Results**

No difficulty was reported during the calibration stage. The procedure was completed by each seated and healthy subject. The mean score was 9.2, with a standard deviation at 0.8.

This result demonstrates three points: (1) young healthy subjects have a strong and accurate perception of the electro-stimulated information; (2) this information is both meaningful and correctly interpreted; and (3) the action resulting from the interpreted information is relevant since the corresponding postural changes decrease the overpressure area.

## III. CAR DRIVING ASSISTANCE

Accurate proprioception at the ankle joint is critical for body orientation and balance control and represents a prerequisite for different functional activities such as walking, running or driving. Indeed, driving a car requires, among other things, accurate foot movements since the drivers control the stopping and the speed of the vehicle with podal actions on brake and accelerator pedals [11]. Normal aging, disease or trauma can cause a loss of sensation in the feet and ankle (e.g., [12-16]), which could prevent drivers from gauging the amount of pressure they're applying to the brake and gas pedals. Impaired ankle proprioception also may be a predisposing factor for chronic ankle instability, balance difficulties, reduced mobility functions, fall, injury and re-injury (e.g. [17-20]). Therefore, it is legitimate to propose that a therapeutic intervention and/or a technical assistance designed to increase proprioceptive acuity at the ankle could be of great interest for performing functional activities such as walking, running or driving a car in the elderly and/or disabled persons.

Along these lines, we developed an original biofeedback system for improving proprioceptive acuity at the ankle joint whose underlying principle consists in supplying the user with supplementary sensory information related to the position of the matching ankle relative to the reference ankle position through the TDU.

The purpose of the present experiment was to assess the performance of this system in young healthy adults.

**Methods**

Subjects

Eight young male healthy university students (mean age: 24.8 ± 1.5 years) were included in this study. They gave their informed consent to the experimental procedure as required by the Helsinki declaration (1964) and the local Ethics Committee, and were naive as to the purpose of the experiment. None of the subjects presented any history of injury, surgery or pathology to either lower extremity that could affect their ability to perform the ankle joint position test.

Task and procedures

Subjects were seated comfortably in a chair with their right and left foot secured to two rotating footplates. The knee joints were flexed at about 110°. Movement was restricted to the ankle in the sagittal plane, with no movement occurring at the hip or knee. The axes of rotation of the footplates were aligned with the axes of rotation of the ankles. Precision linear potentiometers attached on both footplates provided analog voltage signals proportional to the ankles' angles. A handheld press-button allowed recording the matching. Signals from the potentiometers and the press-button were sampled at 100 Hz (12 bit A/D conversion), then processed and stored within the Labview 5.1 data acquisition system.

Subjects were barefoot for all testing, and care was taken to ensure that there were no discernible cues from the sole of the foot before testing. In addition, a panel was placed above the subject's legs to eliminate visual feedback about both ankles position.

The experimenter placed the left reference ankle at a predetermined angle where the position of the foot was maintained by means of a support (e.g., [21]). Subjects therefore did not exert any effort to maintain the position of the left reference ankle, preventing the contribution of effort cues coming from the reference ankle to the sense of position during the test (e.g., [21]). Two matching angular target positions were used: (1) 10° of plantarflexion (P10°) and (2) 10° of dorsiflexion (D10°). These positions were selected to avoid the extremes of the ankle range of motion to minimize additional sensory input from joint and cutaneous receptors (e.g., [22]). Once the left foot had been positioned at the test angle (P10° *vs*. D10°), subject's task was to match its position by voluntary placement of their right leg. When they felt that they had reached the target angular position (i.e., when the right foot was presumably aligned with the left foot), they were asked to press the button held in their right hand, thereby registering the matched position.

This active matching task was performed under two No-TDU and TDU experimental conditions. The No-TDU condition served as a control condition. In the TDU condition, subjects performed the task using a TDU-biofeedback system.

The underlying principle consisted of supplying subjects with supplementary biofeedback about the position of the matching right ankle relative to the reference left ankle position through the TDU. The following coding scheme for the TDU was used (Figure 3):

(1) when both ankles were in a similar angular position within a range of 0.5°, no electrical stimulation was provided in any of the electrodes of the matrix (Figure 3A);

(2) when the position of the matching ankle was determined to be outside the Dead Zone (DZ), electrical stimulation of either the anterior or posterior zone of the matrix (2×6 electrodes) (i.e. stimulation of front and rear portions of the tongue) was provided, depending on whether the matching right ankle was in a too plantarflexed (Figure 3B) or dorsiflexed (Figure 3C) position relative to the reference left ankle, respectively.

Several practice runs were performed prior to the test to ensure that subjects had mastered the relationship between ankle angular positions and lingual stimulations and to gain confidence with the TDU.

Five trials for each target angular position and each experimental condition were performed. The order of presentation of the two targets angular positions (P10° *vs*. D10°) and the two experimental conditions (No-TDU *vs*. TDU) was randomized. Subjects were not given feedback about their performance and errors in the position of the right ankle were not corrected.

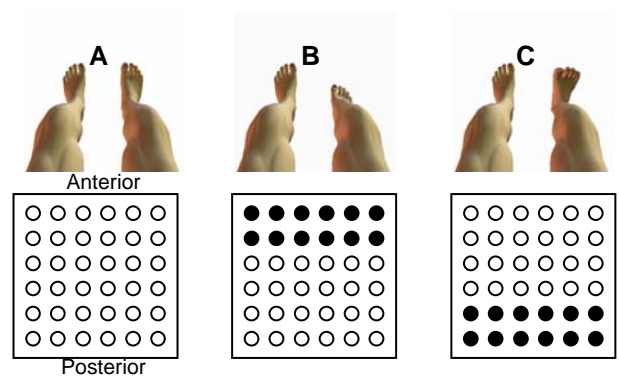

**Figure 3.** Sensory coding schemes for the TDU (*lower panels*) as a function of the position of the matching right ankle relative to the reference left ankle (*upper panels*). Black dots represent activated electrodes. There were 3 possible stimulation patterns of the TDU.
**A**: no electrodes were activated when both ankles are in a similar angular position within a range of 0.5°.
**B**: 12 electrodes (2 × 6) of the anterior zone of the matrix were activated (corresponding to the stimulation of the front portion of the tongue dorsum) when the matching right ankle was in a too plantarflexed position relative to the reference left ankle.
**C**: 12 electrodes (2 × 6) of the posterior zone of the matrix were activated (corresponding to the stimulation of the rear portion of the tongue dorsum) when the matching right ankle was in a too dorsiflexed position relative to the reference left ankle.

Data analysis

Matching performance was quantified using two dependent variables [23].

(1) The absolute error (AE in degree), the absolute value of the difference between the position of the right matching ankle and the position of the left reference ankle, is a measure of the overall accuracy of positioning.

(2) The variable error (VE in degree), the variance around the mean constant error score, is a measure of the variability of the positioning.

Decreased values in AE and VE indicate increased accuracy and consistency of the positioning, respectively (Schmidt, 1988).

The means of the five trials performed in each of the two experimental conditions were used for statistical analyses. Two Conditions (No-TDU *vs.* TDU) × 2 Targets angular positions (P10° *vs.* D10°) analyses of variances (ANOVAs) with repeated measures of both factors were applied to the AE and VE data. Level of significance was set at 0.05.

**Results**

Analysis of the AE showed a main effect of Condition, yielding smaller values in the TDU than No-TDU condition ($F(1,7) = 24.91$, $P < 0.01$, Figure 4A). The ANOVAs showed no main effect of Target angular position, nor any interaction of Condition × Target angular position ($Ps > 0.05$).

Analysis of the VE also showed a main effect of Condition yielding smaller values in the TDU than No-TDU condition ($F(1,7) = 31.37$, $P < 0.001$, Figure 4B). The ANOVAs showed no main effect of Target angular position, nor any interaction of Condition × Target angular position ($Ps > 0.05$).

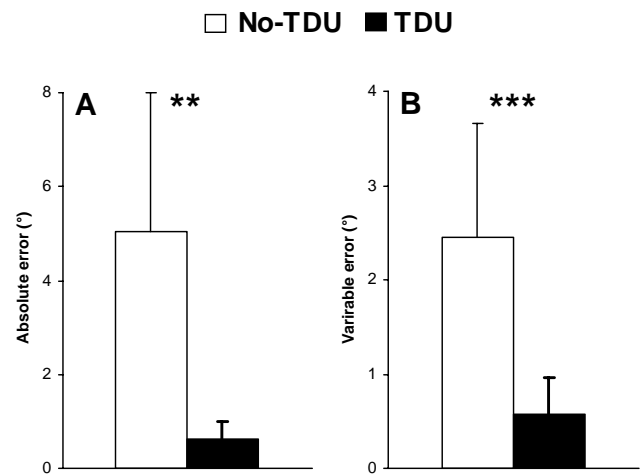

**Figure 4.** Mean and standard deviation for the absolute error (**A**) and the variable error (**B**) for the two No-TDU and TDU conditions. The two experimental conditions are presented with different symbols: No-TDU (*white bars*) and TDU (*black bars*). The significant *P*-values for comparison between No-TDU and TDU conditions also are reported (**: $P<0.01$, ***: $P<0.001$).

These results suggested more accurate and more consistent matching performances, whatever the target angular position, when biofeedback was in use than when it was not.

These results provide evidence that electrotactile stimulation of the tongue can be used to improve ankle proprioceptive acuity.

### IV. FALL PREVENTION

Postural control is a particularly complex system involving various sensory and motor components. Among the sensory inputs relevant to the regulation of postural sway, the importance of somatosensory information from the foot sole is now well established (e.g., [24,25]). Clinically, alteration or loss of somatosensory information from the lower limbs resulting from normal aging or disease (e.g., diabetic peripheral neuropathy) [16,25] is known to impair postural control (e.g., [12,13,26]). Progressive degeneration of sensory inputs from the lower extremities also represents a common clinical finding associated with aging (e.g., [13-16]) and has even been identified as important contributing factor to the occurrence of falls in elderly (e.g., [27,28]). Therefore, it is legitimate to propose that a therapeutic intervention and/or a technical assistance designed to increase somatosensory function of the plantar sole could be of great interest for controlling balance and preventing falls in the elderly and/or disabled persons.

Along these lines, we developed an original biofeedback system for improving postural control whose underlying principle consists in supplying the user with supplementary sensory information related to foot sole pressure distribution through the TDU. The purpose of the present experiment was to assess the performance of this system in young healthy adults.

**Methods**
Subjects

Eight young male healthy university students (mean age: 25.3 ± 3.3 years) were included in this study. They gave their informed consent to the experimental procedure as required by the Helsinki declaration (1964) and the local Ethics Committee, and were naive as to the purpose of the experiment. None of the subjects presented any history of motor problem, neurological disease or vestibular impairment that could affect their ability to perform the postural task.

Task and procedures

Subjects stood barefoot, foot together, their hands hanging at the sides, with their eyes closed. They were asked to sway as little as possible in two No-TDU and TDU experimental conditions. The No-TDU condition served as a control condition. In the TDU condition, subjects performed the postural task using a plantar pressure-based, tongue-placed tactile biofeedback system. A plantar pressure data acquisition system (FSA Inshoe Foot pressure mapping system, Vista Medical Ltd.), consisting of a pair of insoles instrumented with an array of 8×16 pressure sensors per insole (1cm² per sensor), was used. The pressure sensors transduced the magnitude of pressure exerted on each left and right foot sole at each sensor location into the calculation of the positions of the resultant ground reaction force exerted on each left and right foot, referred to as the left and right foot centre of foot pressure, respectively ($CoP_{lf}$ and $CoP_{rf}$). The positions of the resultant CoP were then computed from the left and right foot CoP trajectories through the following relation [29]:

$CoP = CoP_{lf} \times R_{lf} / (R_{lf} + R_{rf}) + CoP_{rf} \times R_{rf} / (R_{rf} + R_{lf})$,

where $R_{lf}$, $R_{rf}$, $CoP_{lf}$, $CoP_{rf}$ are the vertical reaction forces under the left and the right feet, the positions of the CoP of the left and the right feet, respectively.

CoP data were then fed back in real time to the TDU. Note that the TDU was inserted in the oral cavity all over the duration of the experiment, ruling out the possibility the postural improvement observed in the TDU relative to the No-TDU condition to be due to mechanical stabilization of the head in space.

The underlying principle of our biofeedback system was to supply subjects with supplementary information about the position of the CoP relative to a predetermined adjustable "dead zone" (DZ) through the TDU. In the present experiment, antero-posterior and medio-lateral bounds of the DZ were set as the standard deviation of subject's CoP displacements recorded for 10 s preceding each experimental trial.

The following coding scheme for the TDU was used (Figure 5):

(1) when the position of the CoP was determined to be within the DZ, no electrical stimulation was provided in any of the electrodes of the matrix (Figure 5, central panel);

(2) when the position of the CoP was determined to be outside the DZ, electrical stimulation was provided in distinct zones of the matrix, depending on the position of the CoP relative to the DZ (Figure 5, peripheral panels). Specifically, eight different zones located in the front, rear, left, right, front-left, front-right, rear-left, rear-right of the matrix were defined ; the activated zone of the matrix corresponded to the position of the CoP relative to the DZ. For instance, in the case that the CoP was located towards the front of the DZ, a stimulation of the anterior zone of the matrix (i.e. stimulation of the front portion of the tongue) was provided (Figure 5, upper panel).

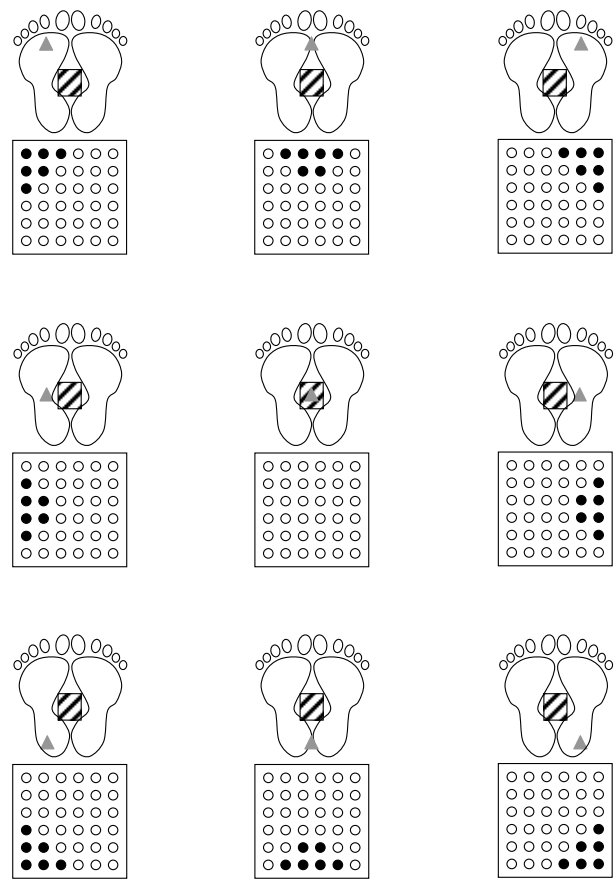

**Figure 5.** Sensory coding schemes for the Tongue Display Unit (TDU) as a function of the position of the centre of foot pressure (CoP) relative to a predetermined dead zone (DZ).

Black triangles, dashed rectangles and black dots represent the positions of the CoP, the predetermined dead zones and activated electrodes, respectively. There were 9 possible stimulation patterns of the TDU.

**Central panel**: no electrodes were activated when the CoP position was determined to be within the DZ.

**Peripheral panels**: 6 electrodes located in the front, rear, left, right, front-left, front-right, rear-left, rear-right zones of the matrix were activated when the CoP positions were determined to be outside the DZ, located towards the front, rear, left, right, front-left, front-right, rear-left, rear-right of the DZ, respectively. These 8 stimulation patterns correspond to the stimulations of the front, rear, left, right, front-left, front-right, rear-left, rear-right portions of the tongue dorsum, respectively.

Several practice runs were performed prior to the test to ensure that subjects had mastered the relationship between the position of the CoP relative to the DZ and lingual stimulations.

A force platform (AMTI model OR6-5-1), which was not a component of the biofeedback system, was used to measure the displacements of the centre of foot pressure (CoP), as a gold-standard system for assessment of balance during quiet standing. Signals from the force platform were sampled at 100

Hz (12 bit A/D conversion) and filtered with a second-order Butterworth filter (10 Hz low-pass cut-off frequency).

Three 30s trials for each experimental condition (No-TDU *vs*. TDU) were performed. The order of presentation of the two experimental conditions was randomized. Subjects were not given feedback about their performance.

Data analysis

Postural control was quantified using two dependent variables.

(1) The surface area (in mm²) covered by the trajectory of the CoP with a 90% confidence interval [30] is a measure of the CoP spatial variability. During quiet standing, increased values in CoP surface area indicate a decreased postural control, whereas decreased values express an increased postural control.

(2) The density histograms of the sway path [31] give a qualitative description of the distribution of the CoP displacements. Distributions were quantified by calculating an average CoP position and the percentage of time the subjects spent in 6 arbitrarily defined concentric circles around the average CoP in radial increments of 2.5 mm. Each of these areas was labelled from area 0-2.5 mm (closest to mean CoP) to area 12.5-15 mm (farthest from mean CoP). During quiet standing, increased values of the percentage of time spent away from the average CoP indicate a narrower sway behaviour, whereas increased values of the percentage of time spent close to the average CoP indicate a wider sway behaviour.

The means of the three trials performed in each of the two experimental conditions were used for statistical analyses. A one-way ANOVA 2 Conditions (No-TDU *vs*. TDU) was applied to the CoP surface area data. A 2 Conditions (No-TDU *vs*. TDU) × 6 Areas (0-2.5 *vs*. 2.5-5 *vs*. 5-7.5 *vs*. 7.5-10 *vs*. 10-12.5 *vs*. 12.5-15 mm) ANOVA with repeated measures of both factors was applied to the density histograms of the sway path data. Level of significance was set at 0.05.

**Results**

Analysis of the surface area covered by the trajectory of the CoP showed a main effect of Condition, yielding a narrower surface area in the TDU than No-TDU condition ($F(1,7) = 25.24$, $P < 0.01$, Figure 6).

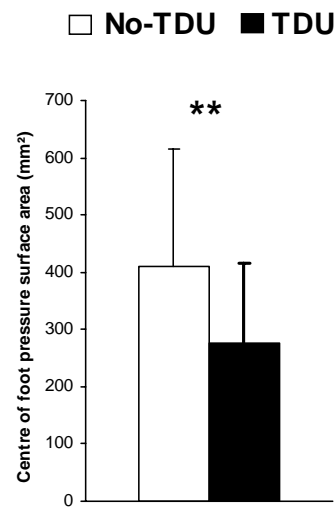

**Figure 6.** Mean and standard deviation of the surface area covered by the trajectory of the centre of foot pressure (CoP) obtained in the two No-TDU and TDU conditions. These experimental conditions are presented with different symbols: No-TDU (*white bars*) and TDU (*black bars*). The significant *P*-values for comparison between No-TDU and TDU conditions also are reported (**: $P < 0.01$).

Analysis of the density histograms of the sway path showed a significant interaction of Condition × Area ($F(5,35) = 7.81$, $P < 0.001$, Fig. 7). The ANOVA also showed main effects of Condition ($F(1,7) = 10.05$, $P < 0.05$) and Area ($F(5,35) = 19.36$, $P < 0.001$). The decomposition of the interaction into its simple main effects showed that subjects spent more time in areas 0-2.5 and 2.5-5 mm in the TDU than No-TDU condition ($P < 0.001$ and $P < 0.05$, respectively), whereas they spent less time in 10-12.5 and 12.5-15 mm in the TDU than No-TDU condition *($Ps < 0.05$)*. In figure 7, condition differences in the dispersion of sway are illustrated by subtracting the percentages of time spent by the subjects in the TDU condition in a given area from those obtained in the No-TDU condition in that same area. Positive values indicate than subjects spent more time in an area in the TDU than No-TDU condition, whereas negative indicate than subjects spent less time in an area in the TDU than No-TDU condition. Clearly, subjects exhibited narrower sway dispersion in the TDU than No-TDU condition.

On the whole, these results suggested that young healthy adults were able to take advantage of an artificial tongue-placed tactile biofeedback to improve their postural control during quiet standing.

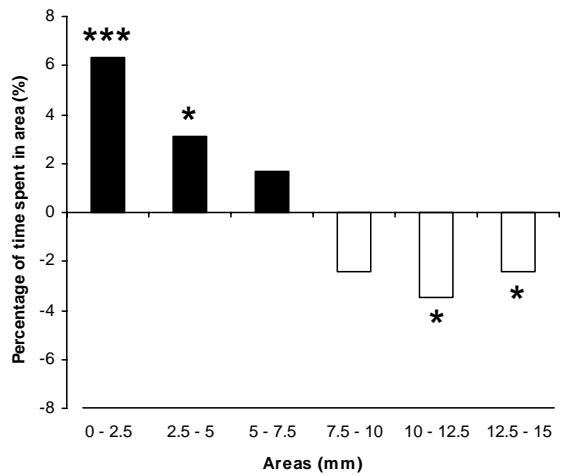

**Figure 7.** Condition differences in the dispersion of sway. For each area, the percentages of time spent by the subjects in the TDU condition were subtracted from those spent in the No-TDU condition. Thus, positive values (*black bars*) indicate that subjects spent more time in an area in the TDU than No-TDU condition, whereas negative values (*white bars*) indicate than subjects spent less time in an area in the TDU than No-TDU condition. The significant *P*-values for comparison between No-TDU and TDU conditions also are reported (*: $P < 0.05$; ***: $P < 0.001$).

## V. DISCUSSION & CONCLUSION

This paper presented three experiments evaluating the feasibility of an artificial tongue-placed tactile biofeedback for perceptual supplementation in the fields of human disability and biomedical engineering.

Proposed applications are: (1) pressure sores prevention in case of spinal cord injuries (persons with paraplegia, or tetraplegia); (2) ankle proprioceptive acuity improvement for driving assistance in older and/or disabled adults; and (3) balance control improvement to prevent fall in older and/or disabled adults.

Overall, the present findings evidence that electrotactile stimulation of the tongue can be used (1) as a part of a device designed to prevent the formation of pressure sores (*experiment 1*), (2) to improve proprioceptive acuity at the ankle (*experiment 2*) and (3) to improve postural control during quiet standing (*experiment 3*). Although these feasibility studies have been conducted in young healthy individuals, we strongly believe that our results could have significant implications in rehabilitative and ergonomical areas, for enhancing/restoring/preserving balance and mobility in individuals with reduced capacities (resulting either from normal aging, trauma or disease) with the aim at ensuring autonomy and safety in occupations of daily living and maximizing quality of life. Along these lines, the effectiveness of our biofeedback system in the three applications presented in this paper is currently being evaluated in paraplegic patients, individuals with somatosensory loss in the feet from diabetic peripheral neuropathy and persons with lower limb amputation, respectively.

Finally, although the current version of our human machine interface (Figure 1A) offers a realistic possibility of practical and cosmetically accepted device for users, the encouraging results evidenced in the present experiments already have led us to improve our system by making it wireless to increase its portability to provide a perspective for the application of this device/technology outside the laboratory framework and to permit its use over long-time period in real-life environment. Indeed, to be acceptable as part of a viable system, this device had to be lightweight, portable, and capable of several hours of continuous operations. The current ribbon TDU system does not meet these requirements yet. Within this context, in addition to the use of a portable, battery-operated ambulatory foot pressure device, we have developed, with the help of Coronis-Systems Company, a wireless radio-controlled version of the 6×6 TDU matrix. This consists in a matrix glued onto the inferior part of an orthodontic retainer including microelectronics, antenna and battery, which can be worn inside the mouth like a dental retainer (Figure 1B). The effectiveness of this wearable device will soon be tested in a daily life situation.